\documentclass[aps,prd,showpacs,nofootinbib,twocolumn,10pt]{revtex4-1}

\usepackage{amsmath}
\usepackage{amsfonts}
\usepackage{amssymb}
\usepackage{graphicx}
\usepackage{color}

\newcommand{\cost}{\cos \theta}

\newcommand{\costs}{\cos^2 \theta}
\newcommand{\sints}{\sin^2 \theta}

\begin{document}
\title{Invariant characterization of the Kerr spacetime: Locating the horizon\newline and measuring the mass and spin of rotating black holes\newline using curvature invariants}
\author{Majd Abdelqader}
\email[]{majd@astro.queensu.ca}
\author{Kayll Lake}
\email[]{lake@astro.queensu.ca}
\affiliation{Department of Physics, Queen's University, Kingston,
Ontario K7L 3N6, Canada}
\date{\today}

\begin{abstract}
We provide an invariant characterization of the physical properties of the Kerr spacetime. We introduce two dimensionless invariants, constructed out of some known curvature invariants, that act as detectors for the event horizon and ergosurface of the Kerr black hole. We also show that the mass and angular momentum can be extracted from local measurements of the curvature invariants, which in the weak field limit could be used to approximate the total angular momentum and mass of a system of merging black holes. Finally, we introduce a dimensionless invariant that gives a local measure of the ``Kerrness'' of the spacetime.
\end{abstract}

\pacs{04.70.Bw, 04.20.-q, 95.30.Sf}

\maketitle

\section{Introduction}

In this paper we present a new approach to analyze and extract physical properties of spacetimes around rotating black holes using curvature scalar invariants. This builds on earlier work \cite{lake1,abdelqader1,Pelavas2001}. However, in this paper we go beyond visualization, and use the invariants, and only the invariants, to locate the horizon and ergosurface, then calculate the mass and angular momentum of the Kerr black hole.

One of the main applications would be in the analysis of numerical relativity simulations. Currently, extracting information about the mass and angular momentum of black holes in numerical simulations requires finding the event horizon of the black hole, calculating the area and angular momentum of the horizon, then using the relationship between the area, mass, and angular momentum in order to calculate the mass \cite{numericalbook}.

In Sec.~\ref{invars}, we state the five curvature invariants on which we base the entire calculations that follow. Furthermore, out of the five curvature invariants, we construct and introduce three new dimensionless invariants. These dimensionless invariants serve as detectors for the horizon and ergosurface of the Kerr black hole.

In Sec.~\ref{sec:global}, we present a global approach to analyze the Kerr spacetime and provide a method to extract its mass and angular momentum. First, we show how some dimensionless invariants can be used to locate the event horizon and ergosurface. Next, the area of these two surfaces can be calculated, and this in turn leads to the angular momentum and mass of the black hole. It is also possible to locate the inner event horizon using one of the dimensionless invariants. Therefore, this could provide an alternative technique to find a region to excise around the singularity to be used in the excision method in numerical relativity.

In Sec.~\ref{sec:local}, we present a procedure to calculate the mass and angular momentum locally. In general, the goal and approach we take in this section are similar to the ones in \cite{Ferrando98, Ferrando2009, GomezLobo2007, Backdahl2010}, but the choice of invariants used to carry out the calculations differs. The invariants we use are of degree 2 (i.e. the contractions involve up to 2 factors of the curvature tensor), and order 3 (i.e. up to the 3rd derivative of the metric tensor). On the other hand, the objects used in the references above are of degree 3 and order 3. We present the procedure for the Kerr metric in general, then derive the simplified expressions for the Schwarzschild black hole (i.e. zero angular momentum), and in the weak field limit with angular momentum. Furthermore, as a by-product of the calculation of the mass and angular momentum locally, it is possible to produce the Boyer-Lindquist coordinates at each point.

The intermediate steps of the procedures are not unique, and it is possible to take a different approach at each step. However, after tedious trial and error, the steps presented here are the ones we found to complete the calculation in the least amount of steps, with the simplest expressions algebraically. However, we show some alternative steps in the Appendix.

In Sec.~\ref{kerrness} we construct the ``Kerrness'' invariant that serves as an invariant local measure of the spacetime deviation from Kerr. It is a dimensionless invariant that ranges from 0 to 1, where a value of 1 indicates a perfect Kerr spacetime locally. A set of invariants was proposed to achieve the same goal in \cite{GomezLobo2012}, but was based on a different choice of invariants.

Finally, in Sec.~\ref{sec:discussion} we discuss the possible applications of the results we present here in the analysis of exact and numerical spacetimes. It is worth noting that the three different procedures presented in Secs.~\ref{sec:global}, \ref{sec:local}, and \ref{kerrness} are complementary but independent. In other words, each procedure can be performed and completed separately. The only common steps between them are the initial ingredients, which are the invariants presented in Sec.~\ref{invars}.

\section{The curvature invariants of Kerr}\label{invars}
Constructing a minimal list of independent curvature invariants that characterize a spacetime is still an active research field \cite{CM91,ZM97}. In Kerr spacetime, all of the Ricci scalars vanish since it is a vacuum solution, and it has been shown that for the Kerr metric there are at most four independent invariants \cite{Coley2009}. Nonetheless, we consider the following seven invariants in this paper
\footnote{The first two invariants are often discussed in the literature as the real and imaginary parts of the complex Weyl invariant. In vacuum solutions such as the Kerr metric, ${C}_{\alpha \beta \gamma \delta} ={R}_{\alpha \beta \gamma \delta}$, where ${R}_{\alpha \beta \gamma \delta}$ is the Riemann tensor. Therefore, in this case $I_1$ equals the Kretschmann scalar. Furthermore, $I_1$ and $I_2$ can be expressed in terms of invariants in the Newman-Penrose formalism, and in the Kerr spacetime $I_1/48=\Re{(\Psi_2)}^2-\Im{(\Psi_2)}^2$, and $I_2/48=-2\Re{(\Psi_2)}\,\Im{(\Psi_2)}$. For a thorough review of the relationship between many curvature invariants in different notations in general see \cite{Cherubini2002}. The differential invariants $I_3$ and $I_4$ were first introduced and analyzed in \cite{karlhede82} and often referred to as the Karlhede invariants, and the differential invariants $I_5$, $I_6$, and $I_7$ were first introduced in \cite{Lake2004}. The gradient fields $k_{\mu}$ and $l_{\mu}$ for the Kerr metric were thoroughly analyzed in \cite{abdelqader1}.}:
\begin{align}
 I_1 &\equiv C_{\alpha \beta \gamma \delta}\;C^{\alpha \beta \gamma \delta} \label{defi1} \;,\\
 \nonumber \\ 
 I_2 &\equiv {C^{*}}_{\alpha \beta \gamma \delta}\;C^{\alpha \beta \gamma \delta}\label{defi2} \;,\\
 \nonumber \\ 
 I_3 &\equiv \nabla_{\mu} C_{\alpha \beta \gamma \delta} \;\nabla^{\mu} C^{\alpha \beta \gamma \delta}\label{defi3}\;,\\
 \nonumber \\ 
 I_4 &\equiv \nabla_{\mu} C_{\alpha \beta \gamma \delta} \;\nabla^{\mu} {C^{*}}^{\; \alpha \beta \gamma \delta}\label{defi4}\;,\\
 \nonumber \\ 
 I_5 &\equiv k_{\mu} k^{\mu}\label{defi5}\;,\\
 \nonumber \\ 
 I_6 &\equiv {l}_{\mu} {l}^{\mu}\label{defi6}\;,
\end{align}
and
\begin{align}
 I_7 &\equiv {k}_{\mu} {l}^{\mu}\label{defi7}\;,\qquad \qquad \qquad \quad\,
\end{align}
where $ {C}_{\alpha \beta \gamma \delta} $ is the Weyl tensor, $ {C^{*}}_{\alpha \beta \gamma \delta} $ its dual, $ k_{\mu} \equiv - \nabla_{\mu}\, I_1\,$, and $ {l}_{\mu} \equiv - \nabla_{\mu}\, I_2 \;$. The explicit expression of these invariants for the Kerr spacetime is given in the Appendix in a compact form.

Only four of the above seven invariants are actually independent. Most importantly, the calculations in the next two sections to locate the horizon, and calculate the mass and angular momentum are carried out with five of the invariants only, without the need for $I_3$ and $I_4$. However, we present them here for completeness, and in order to explore the three syzygies, or constraining equations, in Kerr spacetime between the full set of the seven nonvanishing invariants in Sec.~\ref{kerrness}.

We introduce and define the following three dimensionless invariants constructed entirely out of the five curvature invariants $I_1$, $I_2$, $I_5$, $I_6$, and $I_7$ stated  in Eqs.~(\ref{defi1}), (\ref{defi2}), and (\ref{defi5})--(\ref{defi7})
\footnote{An earlier preprint version of this manuscript required the use of $I_3$ and $I_4$ in the definition of $Q_1$ and $Q_2$. Those earlier definitions are presented here in the Appendix Eqs.~(\ref{q1c}) and (\ref{q2c}). However, we are grateful to Don Page for pointing out two additional syzygies between the seven invariants that we previously missed \cite{Page2015}, which completely eliminated the need for $I_3$ and $I_4$, and simplified the definitions to the ones we present here in Eqs.~(\ref{q1}) and (\ref{q2}).}
\begin{align}
Q_1 & \equiv  \frac{1}{3\sqrt{3}} \frac{({I_1}^2-{I_2}^2)(I_5-I_6)+4\,I_1\,I_2\,I_7}{({I_1}^2+{I_2}^2)^{9/4}}  \; , \label{q1} \; \;\\  
Q_2 & \equiv \frac{1}{27} \frac{I_5\,I_6-{I_7}^2}{({I_1}^2+{I_2}^2)^{5/2}}  \; , \label{q2}
\end{align}
and
\begin{align}
Q_3 \equiv & \frac{1}{6\,\sqrt{3}}\;\frac{I_5+I_6}{({I_1}^2+{I_2}^2)^{5/4}} \; .  \label{q3} \qquad \qquad \qquad
\end{align}

It is worth noting that the term $({I_1}^2+{I_2}^2)$ is positive definite in Kerr spacetime, and it is used in the denominator simply to make the invariants $Q_1$, $Q_2$, and $Q_3$ dimensionless. The significance of $Q_1$ and $Q_2$ will become evident below, as they represent the long sought after invariant detectors for the Kerr black hole ergosurface and event horizon respectively \cite{Moffat,Gass}. $Q_3$ will be used in the calculation of the spin of the black hole locally, but it is redundant since the same calculation could be done with $Q_2$. Nonetheless, we introduce $Q_3$ for the simplicity of the resulting expressions.

\section{Global Approach for Locating the Horizon and Calculating\newline the mass and spin}\label{sec:global}
In this section we present a global approach to calculate the mass and angular momentum in the Kerr spacetime completely based on curvature invariants. To start, we use the invariants to locate two uniquely defined 2D submanifolds. Afterwards, calculating their areas leads us to the mass and spin parameter of the black hole.

\subsection{Locating the horizon and ergosurface}
The two submanifolds we consider here are the outer horizon and outer ergosurface. After evaluating and simplifying $Q_1$, which was defined in Eq.~(\ref{q1}), in Boyer-Lindquist (BL) coordinates we get
\begin{equation}\label{q1bl}
Q_1=\frac{ \left( r^2-a^2\,\costs \right) \left( r^2-2\,m\,r+a^2\,\costs \right)}{m\,{\left( r^2+a^2\,\costs \right)}^{3/2}} \;.
\end{equation}
Therefore, $Q_1$ vanishes when $r=\pm\, a\,\cost$, and at the ergosurfaces where $r=m\pm\sqrt{m^2-a^2\costs}$. Most importantly, $Q_1$ is strictly positive outside the outer ergosurface, vanishes at the ergosurface, then becomes negative as soon as we cross it. Therefore, it is a very convenient invariant to use to detect the ergosurface in Kerr spacetime. Note that $Q_1$ also vanishes at the inner ergo surface, and at $r=\pm\, a\,\cost$. However, these surfaces lie strictly within the outer ergosurface regardless of the values of $m$ and $a$. Therefore, these additional roots of $Q_1$ do not affect its power to detect the outer ergosurface. We should also note that $I_3$ alone has been proposed as a detector for the outer ergosurface, since it does actually vanish at that surface \cite{karlhede82}. However, $I_3$ has many additional roots (nine roots in addition to the outer ergosurface), and some of these roots define surfaces that lie outside the ergosurface, some are inside it, and some actually cross it depending on the values of $m$ and $a$ \cite{Lake2004}. Therefore, it is very difficult to rely on $I_3$ alone as a detector of the outer ergosurface.

Similarly, after evaluating and simplifying $Q_2$, which was defined in Eq.~(\ref{q2}), in BL coordinates we get
\begin{equation}\label{q2bl}
Q_2=\frac{ a^2\,\sints \left( r^2-2\,m\,r+a^2 \right)}{m^2\,{\left( r^2+a^2\,\costs \right)}} \;.
\end{equation}
Therefore, $Q_2$ vanishes on the axis of rotation ($\theta=0$), and on the horizon where $r=m\pm\sqrt{m^2-a^2}$. Most importantly, $Q_2$ is strictly positive outside the outer horizon (except for on the axis of symmetry where it vanishes, but clearly never switches signs crossing the axis), vanishes at the outer horizon, then becomes negative as soon as we cross it. The invariant vanishes again at the inner horizon, and switches signs to positive inside the inner horizon. Therefore, $Q_2$ is a very convenient invariant to use for detecting the horizons in Kerr spacetime. The ability to locate the inner horizon efficiently could be exploited for the excision method in numerical relativity, providing an alternative approach to choose a region around the black hole singularity that is required to be located within the outer event horizon.

An earlier preprint version of this manuscript required the use of $I_3$ and $I_4$ in the definition $Q_2$, which we present in Eq.~(\ref{q2c}). It was recently noted by Page and Shoom \cite{Page2015} that the numerator of $Q_2$ in Eq.~(\ref{q2c}), which dictates its roots, can be written as $I_5\,I_6-{I_7}^2=(k\cdot k)(l\cdot l)-{(k\cdot l)}^2$. In other words, the invariant $Q_2$ vanishes when the two gradient fields $k_{\mu}$ and $l_{\mu}$ are parallel. This led them to propose a generalization of $Q_2$, and introduced an invariant that vanishes on Killing horizons in stationary spacetimes in general \cite{Page2015}.

In the case of zero angular momentum ($a=0$), the Kerr solution reduces to the Schwarzschild metric. Note that in this case, $Q_2$ vanishes everywhere, since $I_6=I_7=0$ in Eq.~(\ref{q2}), or $a=0$ in Eq.~(\ref{q2bl}). However, $Q_1$ does not vanish and it serves as the horizon detector since the ergosurface coincides with the horizon in Schwarzschild spacetime, where it reduces to $Q_1=I_5/{(I_1)}^{5/2}$. However, $I_1$ is positive definite in Schwarzschild, so it is simply the invariant $I_5$ that vanishes and switches signs at the horizons, and only at the horizons. The same is true for the Karlhede invariant $I_3$ which was first observed in \cite{karlhede82}, and the connection between the two invariants is easy to see since $I_5=\tfrac{12}{5}I_1\,I_3$ in Schwarzschild, which is a syzygy that will be explored in detail in Sec.~\ref{kerrness}.

\subsection{Calculating the mass and spin}

In this subsection we present a method to extract the mass and angular momentum of the Kerr black hole based on the curvature invariants presented above. Once the outer horizon and outer ergosurface are found using the invariants $Q_1$ and $Q_2$, their surface areas can be measured directly, based on the geometry of the spacetime. However, the surface area of the outer horizon $\mathcal{A}_{H}$ is also determined by its mass and angular momentum, and it is given by
\begin{equation}\label{horizonA}
\frac{\mathcal{A}_{H}}{8\pi \, m^2}=1+\sqrt{1-A^2}\; ,
\end{equation}
where $A \equiv a/m$ is the dimensionless spin parameter. The surface area of the outer ergosurface $\mathcal{A}_{Ergo}$ is also determined by the mass and angular momentum, and can be evaluated using
\begin{align}\label{ergoA}
\frac{\mathcal{A}_{Ergo}}{8\pi \, m^2}= &\int\limits_0^{\pi} \Bigg\lbrace
 \sqrt{\frac{\sints \left( 1+\sqrt{1-A^2\costs}+A^2\sints \right)}{4 \left(1-A^2\costs \right)}}   \nonumber \\
 &\times \sqrt{\left( 1+\sqrt{1-A^2\costs} \right) } \Bigg\rbrace  \; \mathrm{d}\theta .
\end{align}
The above formula can be expressed in a closed form using elliptical functions, and a thorough analysis of the ergosurface can be found here \cite{Pelavas2001}.

\begin{figure}[t]
\centering
\includegraphics[width=3.0in,angle=0]{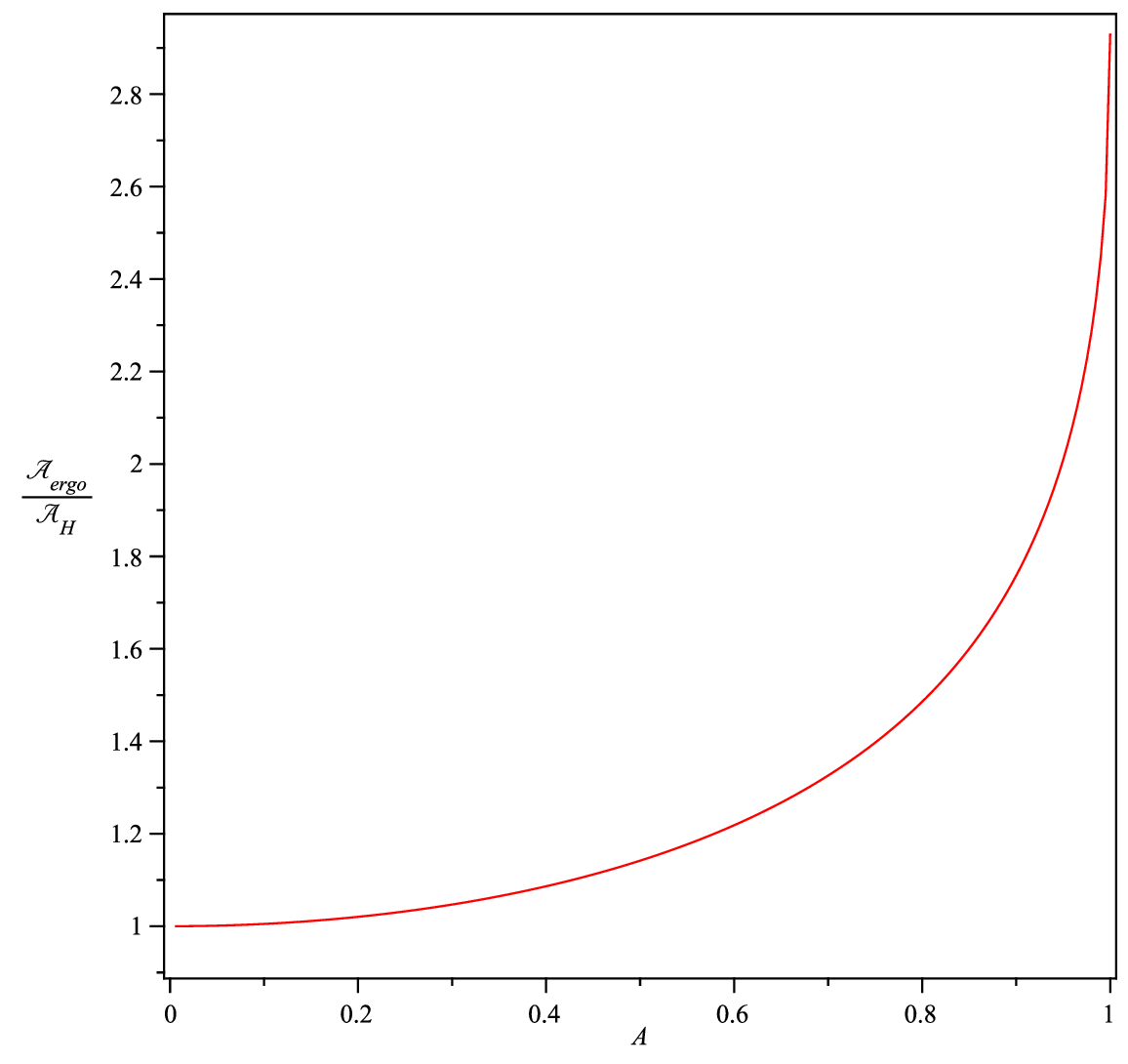}
\caption{\label{fig:area}The ratio between the area of the outer ergosurface to the area of the outer horizon (${\mathcal{A}_{Ergo}}/{\mathcal{A}_{H}}$) for the Kerr black hole as a function of the dimensionless spin parameter $A$.}
\end{figure}

The ratio between the two areas, ${\mathcal{A}_{Ergo}}/{\mathcal{A}_{H}}$, is a one-to-one, strictly increasing function of $A$, which is plotted in Fig.~\ref{fig:area}. Therefore, once $\mathcal{A}_{H}$ and $\mathcal{A}_{Ergo}$ have been measured from the geometry of the spacetime, we can find $A$ directly from the ratio ${\mathcal{A}_{Ergo}}/{\mathcal{A}_{H}}$. Once $A$ is found, we can substitute its value into Eq.~(\ref{horizonA}), and then solve for $m$ to find the value of the mass.

It is worth emphasizing the point that the details of this procedure to extract the mass and spin of the Kerr black hole are not unique, and could be implemented in many ways. In principle, it is possible to construct dimensionless invariants other than $Q_1$ and $Q_2$ such that their roots uniquely define two other surfaces (other than the outer horizon and ergosurface). However, any other two invariants would only be useful for this application if the ratio between the resulting two surface areas is a strictly increasing or decreasing function of the spin parameter $A$. After that, similar steps can be followed as we describe above, where the areas can be measured based on the geometry of the spacetime, then $A$ and $m$ can be extracted from the areas and their ratio.

\section{Local Approach for Calculating the mass and spin}\label{sec:local}

In this section we present another approach to calculate the mass and spin of the Kerr black hole, still within the context of using curvature invariants alone. However, in contrast to the method presented in the previous section, here we calculate $m$ and $A$ based solely on knowing the five invariants ($I_1$, $I_2$, $I_5$, $I_6$, and $I_7$) locally at any point in the spacetime. One of the advantages of this approach is that it does not require locating the black hole or its event horizon. Therefore it could be used to find the total mass and angular momentum of a 2-body system before the merger by applying it in the weak field limit relatively far away from the rotating masses.

We present here the minimal steps needed to carry out the calculations, and include the derivation in the Appendix. To start, we define the dimensionless parameter $ p_1 \equiv a\,\cost / r $, and calculate it using
\begin{align}
p_1 &=-\tan\left[\frac{5}{2} \tan^{-1}\left(\frac{I_2}{I_1}\right) - \tan^{-1}\left(\frac{2\,I_7}{I_5-I_6}\right)  \right]. \label{eqp1}
\end{align}
Next, we introduce a second dimensionless parameter $ p_2 \equiv r/m $, and calculate it using
\begin{equation}\label{eqp2}
p_2 = \frac{2}{1+p_1^2} + Q_1 \frac{\sqrt{1+p_1^2}}{1-p_1^2} \; .
\end{equation}
The mass $m$ can now be found using 
\begin{equation}\label{eqmass}
m_{\text{Kerr}}=\frac{2\sqrt[4]{3}}{ {p_2}^{3/2} \; {(1+{p_1}^2)}^{3/4}\;\sqrt[8]{ {I_1}^2+{I_2}^2} } \; ,
\end{equation}
and the dimensionless spin parameter $A$ can now be found using
\begin{equation}\label{eqA}
A=\sqrt{p_2 \left[1+ Q_3 \sqrt{1+{p_1}^2} - {p_2} \left( 1- {p_1}^2 \right)/2\right] } \quad .
\end{equation}

As it was in the case of the global approach, again the details of this method are not unique. In principle, the steps in Eqs.~(\ref{eqp1})--(\ref{eqA}) used to calculate $p_1$, $p_2$, $m_{\text{kerr}}$, and $A$ could be done in many alternative ways based on the same invariants, and we include some examples in the Appendix. However, the steps we present here were the simplest we could achieve from an aesthetic point of view. 

The resulting equations for $ A $ and $ m$  above are coordinate independent by definition as they were constructed from scalar invariants. However, we can still extract, for example, the BL coordinates as a by-product of the calculations. More specifically, using the definitions of $p_1$ and $ p_2$, we can calculate the BL radius and polar angle:
\begin{align}
\label{rblcoords}
r & =  m\; p_2 \;,
\end{align}
and
\begin{align}
\label{tblcoords}
\cos(\theta)& = p_1\;r/a=  p_1\;p_2/A \; .
\end{align}

\subsection{Special case 1: The Schwarzschild\newline spacetime (a=0) }\label{sbsec:schw}

In the Schwarzschild black hole (i.e. $a=0$), four of the seven invariants we started the calculation with vanish: $I_2$, $I_4$, $I_6$, and $I_7$. This leads to a simple and exact expression for the mass
\begin{align}
{m}_{\text{Schw}} &\equiv \frac{2\sqrt[4]{3}}{ 
{\left(2+\tfrac{I_5}{3\sqrt{3}\,{I_1}^{5/2}}\right)}^{3/2} \;\sqrt[4]{{I_1}} } \; .  \label{eqmass1}
\end{align}
Also the expression for the BL radius in Schwarzschild spacetime (i.e. the areal radius, which we refer to as $\bar{r}$ here) simplifies to
\begin{align}
\bar{r} &\equiv  \frac{2\sqrt[4]{3}}{ 
{\left(2+\tfrac{I_5}{3\sqrt{3}\,{I_1}^{5/2}}\right)}^{1/2} \;\sqrt[4]{{I_1}}} \; .  \label{rschw}
\end{align}
Alternatively, the two equations above can be expressed in terms of the Karlhede invariant $I_3$ instead of $I_5$, by substituting $I_5=\tfrac{12}{5}I_1\,I_3$, which is a syzygy that will be explored in Sec.~\ref{kerrness}.

\subsection{Special case 2: Weak field limit ($r/m \gg 1 $)}\label{sbsec:wfl}
In the weak field limit where $p_2=r/m \gg 1$, we have $(1/p_2)\ll 1$ and $|p_1| \ll 1$, since $p_1=A\cost/(r/m)$. To leading order in $p_1$ we find that $p_1\cong I_2/6I_1$. Also, to leading order in $p_1$, Eq.~(\ref{eqmass}) simplifies to
\begin{align}
{m_{\text{Kerr}}} &\cong  \frac{2\sqrt[4]{3}}{ 
{\left(2+\tfrac{I_5-I_6}{3\sqrt{3}\,{I_1}^{5/2}}\right)}^{3/2} \;\sqrt[4]{{I_1}} } \; .  \label{mwfl}
\end{align}
The expression for the BL radius in Eq.~(\ref{rblcoords}) simplifies to
\begin{align}
r &\cong \frac{2\sqrt[4]{3}}{ 
{\left(2+\tfrac{I_5-I_6}{3\sqrt{3}\,{I_1}^{5/2}}\right)}^{1/2} \;\sqrt[4]{{I_1}}} \; .   \label{rwfl}
\end{align}
Furthermore, the expression for the dimensionless spin parameter $A$ found in Eq.~(\ref{eqA}) simplifies to 
\begin{align}\label{eqAwf}
A  &\cong \sqrt{\left(2+\tfrac{I_5-I_6}{3\sqrt{3}\,{I_1}^{5/2}}\right) \Big( I_6\,I_1- \tfrac{8}{7}I_2\,I_7 \Big) \Big/ \sqrt{27\,{I_1}^7\,}\;  }\; .
\end{align}
The equation above might be the most relevant to the field of numerical relativity, since it can provide a simple and direct way to approximate the total angular momentum of a binary black hole system when evaluated relatively far away from the system before they merge in the weak field limit. Alternatively, the three equations above can be expressed in terms of $I_3$ instead of $I_5-I_6$, and $I_4$ instead of $I_7$, since in the weak field limit $I_5-I_6 \cong \tfrac{12}{5}I_1\,I_3$, and $I_7\cong \tfrac{21}{10} I_1\,I_4$ as a result of the syzygies that will be explored in the next section.

In a recent paper, a procedure to calculate special relativistic linear and angular momentum based on curvature invariants was proposed \cite{Flanagan:2014}. The procedure requires defining two quantities, $M$ and $r$, based on similar invariants we use here. These quantities resemble the zeroth order approximation of ${m}_{\text{Schw}}$ and $\bar{r}$ found in Eqs.~(\ref{eqmass1}) and (\ref{rschw}). However, using Eqs.~(\ref{mwfl}) and (\ref{rwfl}) for $M$ and $r$ instead might improve the accuracy of the procedure proposed in \cite{Flanagan:2014}.

\section{Invariant syzygies and ``Kerrness'' Invariant}\label{kerrness}

Only four of the seven invariants introduced in Sec.~\ref{invars} are independent. There are three syzygies, or constraining equations, between the invariants:
\begin{align}
I_6-I_5+ \tfrac{12}{5} \left(I_1\,I_3-I_2\,I_4 \right)&=0 \;, \label{syzygy} \\
\nonumber \\
I_7- \tfrac{6}{5} \left(I_1\,I_4+I_2\,I_3 \right)&=0 \;, \label{syzygy2}
\end{align}
and
\begin{multline}\label{syzygy3}
4I_1\,I_2\,I_3({I_1}^2-{I_2}^2)({I_3}^2-3{I_4}^2) \\ =I_4(3{I_3}^2-{I_4}^2)({I_1}^4-6{I_1}^2\,{I_2}^2+{I_2}^4) \;.
\end{multline}
The first syzygy, Eq.~(\ref{syzygy}), was discovered by accident, and the other two, Eqs.~(\ref{syzygy2}) and (\ref{syzygy3}), along with a simple way to derive the syzygies were pointed out to us by Don Page \cite{Page2015}, and we include that derivation in the Appendix.

These syzygies can be exploited to construct a geometric invariant measure of the ``Kerrness'' of a spacetime locally. For example, consider the dimensionless invariant $\chi$ defined as
\begin{equation}\label{chi}
{\chi} \equiv \frac{I_6-I_5+ \tfrac{12}{5} \left(I_1\,I_3-I_2\,I_4 \right)}{{\left({I_1}^2+{I_2}^2 \right)}^{5/4}}  \; .
\end{equation}

\begin{figure}[b]
\centering
\includegraphics[width=3.3in,angle=0]{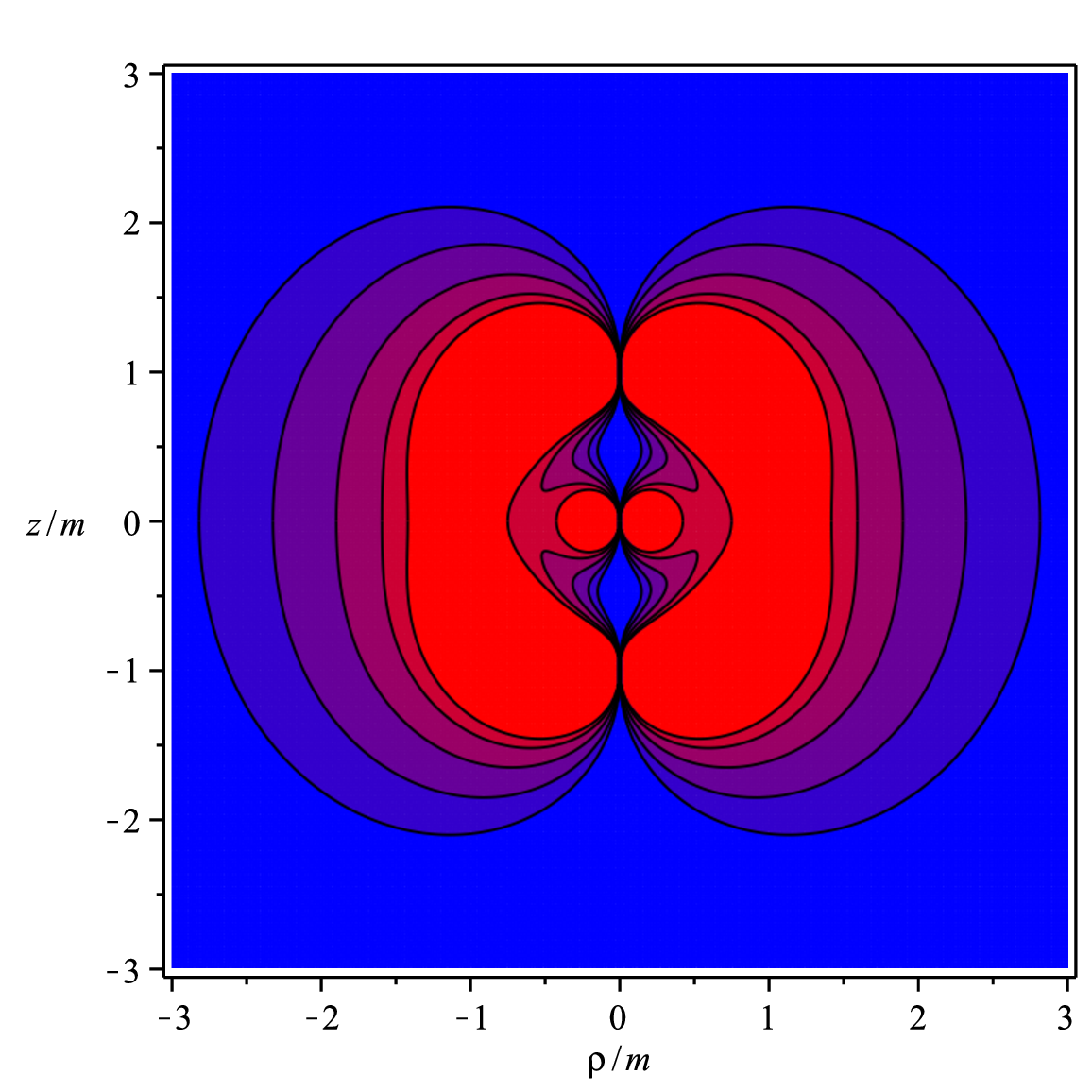}
\caption{\label{fig:kernesscurzon}Contour plot of the Kerrness invariant $K$, which is defined in Eq.~(\ref{K}), in the Curzon-Chazy spacetime. In this plot $s=5$, and the contour levels are 0.01, 0.1, 0.5, 0.85, and 0.96. The lowest contour region (i.e. $0<K<0.01$) is red, and the highest contour region (i.e. $0.96<K<1$) is blue.}
\end{figure}

In Kerr spacetime, evidently $\chi=0$ everywhere as a result of Eq.~(\ref{syzygy}). Furthermore, if we find that $\chi \neq 0$ at some point, this indicates that the local geometry deviates from that of Kerr. However, it is difficult to get an intuitive feel of the scale of this deviation from Kerr directly from the value of $\chi$. Therefore, we construct another dimensionless invariant $K$ based on $\chi$, defined as
\begin{equation}\label{K}
K \equiv e^{-s{\chi}^2} \; ,
\end{equation}
where the constant $s$ is an arbitrary positive number and can be thought of as a sensitivity parameter. By construction, contour levels of $K$ are also contour levels of $\chi$, and both are dimensionless. However, $K$ ranges from 0 to 1. It provides an intuitive measure of how close the spacetime is to the Kerr metric locally, where a value of 1 indicates a perfect Kerr spacetime. The specific value of $K$ is of no real significance, as it is meant to be used in a relative sense, comparing two different points of the spacetime. For example, it can be used to produce what could resemble a heat map of the spacetime, where the regions with the highest values indicate that they are closest to Kerr. We call $K$ the ``Kerrness'' invariant, and propose using it as a spacetime analysis and visualization tool for both exact solutions and numerical relativity simulations.

As an example, we show in Fig.~\ref{fig:kernesscurzon} the contour plot of $K$ for the Curzon-Chazy spacetime \cite{Curzon,Chazy}. This spacetime is believed to be a vacuum spacetime containing a nonrotating singular ring. For a detailed analysis of the Curzon-Chazy solution and its curvature invariants see \cite{abdelqader2} and references within. 

\section{Discussion and Conclusion}\label{sec:discussion}

We have presented an invariant characterization of the Kerr spacetime. The physical properties of a rotating black hole such as its mass, angular momentum, event horizon, and ergosurface can be defined and described in a coordinate-independent and observer-independent formulation based only on curvature invariants. 

The dimensionless invariants introduced in Sec.~\ref{invars} serve as convenient detectors of the black hole's event horizons and ergosurfaces. This can provide an alternative approach to locate black holes in numerical relativity, and choose an appropriate region to excise around the singularity in the excision method by locating the inner horizon. 

Furthermore, the area of the outer ergosurface, along with the area of the outer horizon, provide an alternative method to calculate the mass and angular momentum of rotating black holes as explained in Sec.~\ref{sec:global}. In contrast with current methods used in numerical relativity analysis, this method is completely path independent, and does not require finding the Killing field on the horizon.

Another approach was presented in Sec.~\ref{sec:local}, where the mass and spin parameter can be calculated locally. This procedure could be useful in numerical relativity as well, but will only be reliable in the regions of the spacetime dominated a single Kerr black hole. In numerical simulations of black hole mergers, we expect this procedure to be most reliable near each black hole, where the gravitational field is dominated by one of them, and very far from the system, where the spacetime asymptotically approaches Kerr again but representing the combined mass and angular momentum of the system. However, in the region in between the black holes, the calculations could produce unphysical values of the spin parameter outside the range of 0 to 1, or imaginary numbers. Nonetheless, one of the main advantages of this method is that it does not require finding the horizon, or calculating any surface areas. Therefore, when it is applied in the weak field limit in numerical simulations of black hole mergers, we expect it to give reliable results for the combined mass and angular momentum of the system even before merging.

The Kerrness invariant was introduced in Sec.~\ref{kerrness}, and it can be used in the analysis of exact and numerical spacetimes. For example, it can be used to produce what could resemble a heat map of the spacetime, where the regions with the highest values indicate that this region is closest to Kerr. This provides an invariant and intuitive method to compare and visualize spacetimes. Furthermore, it can be used in combination with local calculations of the mass and spin, as it can indicate the regions where the calculations can be trusted.

\begin{acknowledgments}
The authors would like to thank Don Page for his constructive review of an early version of this manuscript and pointing out two of the syzygies, which greatly improved and simplified the presentation of the results. M.~A. would like to thank Alan Coley and the Department of Mathematics and Statistics at Dalhousie University for their hospitality, where part of this research was conducted while visiting. M.~A. would like to thank Alan Coley, Robert Owen, and Nathan Deg for helpful discussions. K.~L. would like to thank Eric Poisson and Matt Visser for discussions at a very early stage of this work. This work was supported in part by a grant (to K.~L.) from the Natural Sciences and Engineering Research Council of Canada. Portions of this work were made possible by use of \textit{GRTensorII} \cite{grt}.
\end{acknowledgments}

\appendix*
\section{DERIVATION}

In BL coordinates and using natural units ($G=c=1$), the Kerr metric can be expressed as \cite{kerrbook}
\begin{align}
ds^2=&-\left[ 1-\frac{2mr}{r^2+a^2 \cos^2\theta} \right]dt^2- \frac{4mr\,a\,\sin^2 \theta }{r^2+a^2 \cos^2\theta}\, dt\, d\phi \nonumber \\
 &+\left[\frac{r^2+a^2 \cos^2\theta}{r^2-2mr+a^2} \right]dr^2+(r^2+a^2\cos^2\theta) \, d\theta^2 \nonumber \\
 &+\left[ r^2+a^2+\frac{2mr\,a^2\,\sin^2 \theta }{r^2+a^2 \cos^2\theta} \right]\sin^2\theta \; d\phi^2 ,\label{kerrmetric}
\end{align}
where $m$ is the mass, and $a=J/m$ is the angular moment per unit mass, or spin parameter. The seven invariants we start with in Sec.~\ref{invars} can be written explicitly--for the Kerr spacetime--in a compact form  as the real and imaginary parts of three complex invariants, and one purely real invariant as follows
\begin{align}
\mathcal{W}_1 &\equiv I_1+i\,I_2=\frac{48\,m^2}{{\left( r+i\,a\,\cost \right)}^{6}}\; , \label{W1}
\end{align}
\begin{align}
\mathcal{W}_2 &\equiv I_3+i\,I_4 \nonumber \\
 &=\frac{-720\,m^2 \left(r^2-2\,r\,m+a^2\costs \right)}{\left( r^2+a^2\,\costs \right){\left( r+i\,a\,\cost \right)}^{8}} \; , \label{W2} \\
 & & \nonumber \\
\mathcal{W}_3 &\equiv \nabla_{\mu}\mathcal{W}_1\,\nabla^{\mu}\mathcal{W}_1 \nonumber \\
&=I_5-I_6+i\,2\,I_7  \nonumber \\
&=\frac{2^{10}\,3^{4}\,m^4 \left(r^2-2\,r\,m+a^2\costs \right)}{\left( r^2+a^2\,\costs \right){\left( r+i\,a\,\cost \right)}^{14}} \; ,\label{W3}
\end{align}
and
\begin{align}
\mathcal{W}_4 &\equiv \nabla_{\mu}\mathcal{W}_1\,\nabla^{\mu}\overline{\mathcal{W}_1} \nonumber \\
&= I_5+I_6  \nonumber \\
&=\frac{2^{10}\,3^{4}\,m^4 \left(r^2-2\,r\,m+2\,a^2-a^2\costs \right)}{{\left( r^2+a^2\,\costs \right)}^{8}} \; , \label{W4} 
\end{align}
where $\overline{\mathcal{W}_i}$ is the complex conjugate of $\mathcal{W}_i$.

Note that the dimensionless invariants $Q_1$, $Q_2$, and $Q_3$, which were defined in Eqs.~(\ref{q1})--(\ref{q3}), can be written in a compact form using the complex invariants above as
\begin{align}\label{q1b}
Q_1=\frac{ \Re{\left({(\mathcal{W}_1)}^2\,\overline{\mathcal{W}_3} \right)}}{3\sqrt{3}\;{|\mathcal{W}_1|}^{9/2} } \; ,
\end{align}
\begin{align}\label{q2b}
Q_2=\frac{ {(\mathcal{W}_4)}^2 - {|\mathcal{W}_3|}^2}{108\;{|\mathcal{W}_1|}^{5} } \; ,
\end{align}
and
\begin{align}\label{q3b}
Q_3=\frac{ \mathcal{W}_4 }{6\sqrt{3}\;{|\mathcal{W}_1|}^{5/2} } \; .
\end{align}

In order to find the parameter $p_1 \equiv a\cost/r$, we calculate and simplify the complex invariant $\mathcal{W}_5$
\begin{flalign}\label{W5}
\mathcal{W}_5 &\equiv {\left( \mathcal{W}_1 \right)}^{5/2} \;  \overline{\mathcal{W}_3}  & \nonumber \\
&=\frac{\pm\; 2^{20}\,3^{13/2}\,m^9\,r \left(r^2-2\,r\,m+a^2\costs \right)}{{\left( r^2+a^2\,\costs \right)}^{16}}  & \nonumber \\
&\times \left( 1-i\,p_1 \right) \; . &
\end{flalign}
Therefore, $p_1=-\tan(\varphi_5)$, where $\varphi_i$ is the argument of $\mathcal{W}_i$. However, from the definition of $\mathcal{W}_5$, we get $\varphi_5=\tfrac{5}{2}\varphi_1-\varphi_3$. This leads to
\begin{align}
p_1 &=-\tan\left[\frac{5}{2}\varphi_1-\varphi_3 \right] \label{derp1} \; ,
\end{align}
which leads to Eq.~(\ref{eqp1}) after writing the arguments in terms of the invariants since $\varphi_1=\tan^{-1}\left(\frac{I_2}{I_1}\right)$ and $\varphi_3=\tan^{-1}\left(\frac{2\,I_7}{I_5-I_6}\right)$.

Next we substitute $ a\cost=r\,p_1$, and introduce a second dimensionless parameter $ p_2 \equiv r/m$ into $Q_1$, which was evaluated in Eq.~(\ref{q1}); then we solve for $p_2$, and we obtain Eq.~(\ref{eqp2}).

Next, we evaluate and simplify $Q_3$, which was defined in Eq.~(\ref{q3}). We find 
\begin{equation}\label{derq3}
Q_3  = \frac{r^2-2\,r\,m+2\,a^2-a^2\costs}{2\,m\,\sqrt{r^2+a^2\,\costs} } \; ,
\end{equation}
and also observe that
\begin{equation}\label{dermass}
{I_1}^2+{I_2}^2=|\mathcal{W}_1|^2=\frac{2^8\,3^2\,m^4}{{\left( r^2+a^2\,\costs \right)}^6} \; .
\end{equation}

Finally , we substitute $ r=m\,p_2$, and $\cost=r\,p_1/a=p_1\,p_2/A$, into Eq.~(\ref{dermass}), then solve for $m$, which produces Eq.~(\ref{eqmass}). We do the same for (\ref{derq3}), then solve for $A$, which produces Eq.~(\ref{eqA}). This completes the derivation of the general case.

In the Schwarzschild case ($a=0$), four of the seven invariants vanish: $I_2=0$, $I_4=0$, $I_6=0$, and $I_7=0$. Therefore, the formulas simplify significantly, and we get $p_1=0$ which is clear from the definition of $p_1$, or by using Eq.~(\ref{eqp1}) with $I_2=0$ and $I_7=0$. Since $p_1=0$, Eq.~(\ref{eqp2}) simplifies to 
\begin{equation}\label{derp2sch}
p_{2-\text{Schw}}=2+ \frac{ I_5 }{3\sqrt{3}\;{I_1}^{5/2}}\; .
\end{equation}
Furthermore, substituting $p_{2-\text{Schw}}$ from Eq.~(\ref{derp2sch}) into Eq.~(\ref{eqmass}) produces Eq.~(\ref{eqmass1}). Multiplying  $p_{2-\text{Schw}}$ from Eq.~(\ref{derp2sch}) by the mass from Eq.~(\ref{eqmass1}) leads to Eq.~(\ref{rschw}).

In the weak field limit where $p_2=r/m \gg 1$, we have $|p_1| =|A\cost/(r/m)| \ll 1$. To leading order in $p_1$ we have $I_2/I_1\cong-6\,p_1$. Therefore, $p_1\cong -I_2/6I_1$, and Eq.~(\ref{eqp2}) simplifies to
\begin{equation}\label{derp2wfl}
p_2 \cong 2+ \frac{ I_5-I_6 }{3\sqrt{3}\;{I_1}^{5/2}}\; .
\end{equation}
Furthermore, substituting $p_2$ from Eq.~(\ref{derp2wfl}) into Eq.~(\ref{eqmass}) produces Eq.~(\ref{mwfl}) to leading order in $p_1$. Multiplying  $p_2$ from Eq.~(\ref{derp2wfl}) by the mass from Eq.~(\ref{mwfl}) leads to Eq.~(\ref{rwfl}).

However, we need to proceed with caution for the weak field limit of $A$. We need to include up to second order terms in $p_1$ (i.e. keep ${I_2}^2/{I_1}^2$ terms) in the intermediate calculations, and cancel them in the final expression. Up to leading order in $p_1$, Eq.~(\ref{eqA}) simplifies to
\begin{align}
A &\cong  \sqrt{\left(2+\tfrac{I_5-I_6}{3\sqrt{3}\,{I_1}^{5/2}}\right)  \Big/ \sqrt{27\,{I_1}^9\,}\;  } \nonumber \\
 &\times \sqrt{\Big({I_2}^2\,I_5 +{I_1}^2\,I_6- 2\,I_1\,I_2\,I_7 \Big)} \;. \label{derAwf}
\end{align}
We can further simplify the above equation by noting that up to leading order in $p_1$ in the weak field limit we have $I_2\,I_5/I_1\,I_7\cong 6/7$, which leads to Eq.~(\ref{eqAwf}).

The syzygy presented in Eq.~(\ref{syzygy}) was discovered accidentally. However, Don Page pointed out a simple derivation which led to discovering two additional independent syzygies of the Kerr spacetime \cite{Page2015}, and we include their derivation here. Note that from the definition of the complex invariants Eqs.~(\ref{W1})--(\ref{W3}), we have
\begin{align}\label{dersyz1}
\mathcal{W}_3=\tfrac{12}{5}\mathcal{W}_1\,\mathcal{W}_2 \; .
\end{align}
The real and imaginary parts of the equation above are the syzygies in Eqs.~(\ref{syzygy}) and (\ref{syzygy2}) respectively. Furthermore, we have
\begin{align}\label{dersyz2}
{\left( \mathcal{W}_1 \right)}^{4} \, {\left( \overline{\mathcal{W}_2} \right)}^{3}=\frac{720^3\,48^4\,m^{14} \left(r^2-2\,r\,m+a^2\costs \right)}{{\left( r^2+a^2\,\costs \right)}^{27}} \;,
\end{align}
which is a purely real expression. Therefore, 
\begin{align}\label{dersyz3}
\Im{\left( {\left( \mathcal{W}_1 \right)}^{4} \, {\left( \overline{\mathcal{W}_2} \right)}^{3} \right) }= 0 \;,
\end{align}
which is the syzygy in Eq.~(\ref{syzygy3}). It is possible to produce more syzygies, but they would not be independent from the ones mentioned. For example, we can produce another syzygy between the invariants $I_1$, $I_2$, $I_5$, $I_6$, and $I_7$ directly by noting that the expression
\begin{align}\label{dersyz4}
{\left( \mathcal{W}_1 \right)}^{7} \, {\left( \overline{\mathcal{W}_3} \right)}^{3}= \frac{2^{58}\,3^{19}\,m^{26} \left(r^2-2\,r\,m+a^2\costs \right)}{{\left( r^2+a^2\,\costs \right)}^{45}} \;,
\end{align} 
is a purely real expression. Therefore,
\begin{align}
\Im{\left( {\left( \mathcal{W}_1 \right)}^{7} \, {\left( \overline{\mathcal{W}_3} \right)}^{3} \right) }=0 \; ,
\end{align}
which produces yet another syzygy that can be written in a compact form as $7\,\varphi_1=3\,\varphi_3$.

Finally, we would like to reiterate the fact that the steps to calculate $p_1$, $p_2$, $m$, and $A$ can be done in many alternative ways. For example, using the syzygy expressed in Eq.~(\ref{dersyz1}), we can replace the use of $I_5$, $I_6$, and $I_7$ by the invariants $I_3$ and $I_4$ in the definition of the dimensionless invariants $Q_1$ and $Q_2$. By substituting Eq.~(\ref{dersyz1}) into Eqs.~(\ref{q1b}) and (\ref{q2b}) we obtain
\begin{align}\label{q1c}
Q_1=\frac{4\, \Re{\left(\mathcal{W}_1\,\overline{\mathcal{W}_2} \right)}}{5\sqrt{3}\;{|\mathcal{W}_1|}^{5/2} }=\frac{4}{5\sqrt{3}}\frac{I_1\,I_3+I_2\,I_4}{({I_1}^2+{I_2}^2)^{5/4}} \; ,
\end{align}
and
\begin{align}\label{q2c}
Q_2 &= \frac{ {(\mathcal{W}_4)}^2 - {\tfrac{12}{5}|\mathcal{W}_1\,\mathcal{W}_2|}^2}{108\;{|\mathcal{W}_1|}^{5} } \nonumber \\
 &=\frac{\left(I_5 +I_6 \right)^2-(12/5)^2 \left( {I_1}^2 +{I_2}^2 \right) \left( {I_3}^2 +{I_4}^2 \right) }{108\left( {I_1}^2 +{I_2}^2 \right)^{5/2}} \; .
\end{align}
The mass can also be found by substituting $r=m\,p_2$, and $\cost=p_1\,p_2/A$, into ${|\mathcal{W}_3|}^2$ instead of ${|\mathcal{W}_1|}^2$, then solving for $m$. Similarly, $A$ can be found by performing the same substitution into $Q_2$ instead of $Q_3$, then solving for $A$.

\end{document}